\documentclass{aa}
\usepackage{graphicx}

\begin{document}
 \title{Significant evolution of the stellar mass-metallicity relation since $z$$\sim$0.65} 

  \titlerunning{The evolution of stellar mass-metallicity relation}

   \author{ Y. C. Liang\inst{1,2}, F. Hammer \inst{2} and
          H. Flores\inst{2}
       }

   \authorrunning{Liang et al.}
   
   \offprints{ Y. C. Liang: ycliang@bao.ac.cn     }
   
   \institute{
  $^1$National Astronomical Observatories, Chinese Academy of Sciences, 
 20A Datun Road, Chaoyang District, Beijing 100012, P.R. China,
  email: ycliang@bao.ac.cn;\\
  $^2$GEPI, Observatoire de Paris-Meudon, 92195 Meudon, France
}

 \date{Received; accepted}

\abstract{ We present the stellar mass-metallicity relation for 34 0.4 $<$ $z$
$<$ 1 galaxies selected from CFRS and Marano fields, and compare it to those
derived from three local samples of galaxies (NFGS, KISS  and  SDSS). Our metal
abundance estimates account for extinction effects, as estimated from
IR/optical ratios and Balmer line ratios.   All three comparisons show that the
intermediate mass galaxies at  $z$$\sim$ 0.65 are more metal-deficient by 0.3
dex at a given $M_K$ or stellar mass relative to $z$=0. We find no evidence
that this discrepancy could be related to different methods used to derive mass
and metallicity. Assuming a closed box model predicts  a gas fraction converted
into stars of 20-25\% since $z$$\sim$ 0.65, if the gas fraction is 10-20\% in
present-day galaxies with intermediate masses. This result is in excellent
agreement with previous findings that most of the decline of the cosmic star
formation density is related to the population of intermediate mass galaxies,
which is composed of 75\% spirals today. We find no evidence for a change of
the slope of the $M_{star}$-Z relation from $z$$\sim$0.65 to $z$=0 within the
intermediate mass range (10.5 $< $ log($M_{\star}$) $<$ 11.5).

}

 \maketitle

\section{Introduction}

The origin of the decline of star formation density over the last 8 Gyrs is
still a matter of debate. Studies based on spectral energy distributions (SEDs)
of galaxies  predict that 30\% to 50\% of the mass locked in stars in present
day galaxies condensed into stars at $z$$<$1 (Dickinson et al. 2003; Pozzetti
et al. 2003; Drory et al. 2004; Bell 2004). This result is supported by the
integrations of the star formation density, especially when it accounts for
infrared light (Flores et al. 1999). Indeed, the rapid density evolution of
luminous infrared galaxies (LIRGs) suffices in itself to account for the
formation of $\sim$40\% of the total stellar mass found in intermediate massive
galaxies (from 2$\times$$10^{10}$ to 2$\times$$10^{11}$ $M_{\odot}$, Hammer et
al. 2005). As opposed to the idea of galaxy ``downsizing'' (Cowie et al. 1996)
-- strong evolution only in the faint blue dwarf population  -- there is a
growing consensus that most of the decline of the star formation density  is
indeed related to intermediate-mass galaxies (Hammer et al. 2005; Bell et al.
2005). Recall that those account for most of the total stellar mass today, and
are crowding the deep redshift surveys at intermediate redshifts. Today,
intermediate mass galaxies are mainly spirals, which has led Hammer et al.
(2005) to claim that most of the recent mass evolution has occurred in
progenitors of spirals.

An important test to confirm this simple scenario for the decline of star
formation  in the universe is to study the relation between metal abundances
and stellar masses. This can probe either the galaxy ``downsizing'' scenario
(in which case no evolution is expected at intermediate masses) or the findings
of Hammer et al. (2005). Previous studies of gas metal content in distant
galaxies have related oxygen abundances to absolute B band luminosities 
(Kobulnicky \& Zaritsky 1999, Kobulnicky et al. 2003, Kobulnicky \& Kewley
2004, Liang et al. 2004,  Maier et al. 2004, Lilly et al. 2003, Hammer et al.
2005 etc.). Most of these studies led to the conclusion for evolution of the
$M_{B}$-(O/H) relation, although there are some differences in the magnitude of
the evolution. The oxygen abundances are derived from
$R_{23}$=\ion{[O}{ii]}$\lambda$3727/$H\beta$ +
\ion{[O}{iii]}$\lambda$4959,5007/$H\beta$, it is vital to properly estimate the
\ion{[O}{ii]}$\lambda$3727/$H\beta$ ratio, which thus imply to derive as
accurate as possible the extinction correction factor. Liang et al. (2004) and
Hammer et al. (2005) used very deep spectroscopy of distant galaxies at VLT  to
derive extinction corrected $R_{23}$ values, on the basis of a check of the
extinction correction, using both Balmer line ratio (after a proper correction
for the underlying absorption) and the IR/visible light ratio. They claimed an
evolution of the $M_{B}$-(O/H) relation by $\sim$ 0.3 dex at $z$=0.7, if this
evolution is associated with a dimming of the metal abundance at high redshift.
However, because blue light is easily affected by star formation events, it is
possible that a significant part of the evolution is caused by a brightening
(of 2.5 magnitudes) in the blue band.

The near-IR K-band luminosity is less affected by star formation and by dust
than the B-band luminosity, and is more directly related to the stellar mass
(Charlot 1998; Bell\& de Jong 2000). Therefore, in this study, we have derived
the mass(or $M_K$)-metallicity relations for a sample of intermediate-$z$
galaxies, and compared them with those for local galaxies. The paper is
organized as follows: Sect.2 describes the observational data; Sect.3 describes
the methodology adopted to ensure a fair comparison with local galaxies;
results and analyses are given in Sect.4;  and Sect.5 gives the  conclusion.
Throughout this paper, a cosmological model with  $H_0$=70 km s$^{-1}$
Mpc$^{-1}$, $\Omega _M$=0.3 and $\Omega _\Lambda =0.7$ has been adopted. The AB
magnitude system was used.

\section{Observational data}

The VLT/FORS observations of our intermediate-$z$ galaxies have been described
in Hammer et al. (2005) and Liang et al. (2004).  In addition, we have measured
stellar masses for seven additional objects which possess estimated metal
abundances. In total (see Table 1), 34 star-forming galaxies  (non-AGNs from
the diagnostic diagram) with $z$$>$0.4 are selected from the
Canada-France-Redshift Survey (CFRS) and Marano fields (the Ultra Deep Fields)
with both of their oxygen abundances and K-band absolute magnitudes available,
including luminous IR galaxies (LIRGs), and starbursts. 

The metallicities of the sample galaxies were estimated from the 
extinction-corrected $R_{23}$ parameters to be converted to oxygen abundances
by using the calibration derived from SDSS data by Tremonti et al. (2004).  We
must keep in mind that the only way to derive O/H metal abundances of galaxies
with sufficient accuracy is by deriving the extinction and the underlying
Balmer absorption properly, on the basis of good-S/N spectra of moderate
resolution. The high quality VLT/FORS spectroscopic data help us to obtain
reliable oxygen abundances of the sample galaxies. 
As discussed in Liang et al. (2004), the corresponding error budget of emission
line flux is deduced by a quadratic addition of three independent errors: the
use of stellar templates to fit stellar absorption lines and continuum; the
differences among independent measurements performed by Y.C. Liang, H. Flores
and F. Hammer; and the Poisson noises from both sky and object spectra.
Therefore, most of the uncertainties of 12+log(O/H) range from 0.03 to
0.30\,dex with the average value of 0.10\,dex and the typical value of
0.09\,dex (omitting one object with big error bar which was not used in the
actual statistics, see Table\,1).

The stellar masses of these galaxies were estimated from K-band magnitudes. We
have chosen a conservative approach, which assumes that  $M/L_K$  depends on
the rest-frame B-V color following the   relation derived by Bell et al. (2003)
(see Hammer et al. 2005 for more details). 
Sources of error include uncertainties from magnitude errors, systematic   
uncertainties in stellar $M/L$ ratio from dust and bursts of Star Formation
(SF),  and Poisson errors. The typical uncertainty of the estimated $M_K$ is
about $\pm$0.3\,dex.  The typical error bars of the estimates of the sample
galaxies have been plotted on the left-top on Fig.1.

\section{Comparison to three samples of local galaxies: NFGS, KISS and SDSS data}

We have considered three local samples for comparison with our intermediate-$z$
galaxies. They are the Nearby Field Galaxy Survey (NFGS, Jansen et al. 2000a,b), 
which is a sample of local spirals, the  KPNO International Spectroscopic Survey
(KISS,  Salzer et al. 2005), which is a sample  of local starburst galaxies
selected by their $H\alpha$ emission line, and the Sloan Digital Sky Survey (SDSS,
Tremonti et al. 2004). Because most local galaxies with emission lines are spirals,
it is expected that the NFGS and SDSS should provide essentially a similar
luminosity vs. metal relation. In the following we describe in details how we have
carried out the local data to ensure that the estimates of both O/H and  $M_K$ (or
stellar masses) have done in a similar manner for local and for distant samples. A
brief description of the comparison is presented in Fig.\,1-3.

\subsection{Comparison to local spirals from the NFGS}

Jansen et al. (2000a,b) have published  the emission line fluxes for their 198
nearby field galaxies, including \ion{[O}{ii]}, \ion{[O}{iii]}, $H\beta$, $H\alpha$ etc. This
makes it possible for us to directly estimate the dust extinction of the galaxies
by assuming the same extinction law as we used in Liang et al. (2004), and then to
derive their oxygen abundances. We use the SDSS calibration from Tremonti et al.
(2004)  to convert $R_{23}$ in oxygen abundances for all sources.

The K-band magnitudes of the NFGS sample were derived from 2MASS (Extended Source
Catalog, see http://irsa.ipac.caltech.edu/applications/Gator/).  Four magnitudes
were obtained: k-m-k20fe (K 20mag/sq.), the isophotal fiducial ell. ap. magnitude;
k-m-fe, the K fiducial Kron ell. mag aperture magnitude; k-m-ext, the K mag from
fit extrapolation; and k-m-e, the K Kron elliptical aperture magnitude. Here we
just consider the k-m-ext mag (the K mag from fit extrapolation), which was
assumed to better trace the galactic luminosity. 

Figure\,1 plots the $M_K$-metallicity relations for our sample galaxies, compared
with the local sample from NFGS (E and S0 galaxies have been removed here).  For
distant galaxies, the median oxygen abundance is 12+log(O/H)=8.67 at the median
$M_K$=$-$21.74, which is $\sim$0.3 dex lower (i.e. $\sim$50\% more metal deficient)
than the local ones. Recall that the discrepancy is significantly larger than
typical uncertainties.

\subsection{Comparison to local starbursts from the KISS } 
      
Salzer et al. (2005) derived the luminosity-metallicity relation in both optical
and the near IR (J,H,K bands from 2MASS) for the KISS data.  Oxygen abundances
based on the SDSS calibration were used here.

Figure\,2  reveals a similar discrepancy as seen in Figure\,1 when comparing the
distant to the nearby galaxies.  However, we notice that the local starbursts show
a wide dispersion around the linear least-square fit (also see Salzer et al. 2005)
and hence that there is a significant overlap between the local and distant
starbursts.

\subsection{Comparison to  SDSS galaxies}

SDSS provides a powerful database to study the formation and evolution of
galaxies. Tremonti et al. (2004) have published the metallicities and stellar
masses for $\sim$53400 galaxies.  Bell et al. (2003) re-derived the stellar
masses of a sub-sample of the SDSS galaxies using K band data from 2MASS,  and we
choose to use them to be consistent with our derived stellar masses.   Notice that
the estimates obtained by Bell et al.  provides stellar masses slightly smaller
than Tremonti et al.'s estimates with the discrepancy increasing at the high mass
end (it is $\sim$0.17 dex at $log(M_{\star}$)=10.5). Recall that a Kroupa (2001) IMF was
used in Tremonti et al. (2004), whereas Bell et al. (2003) adopted the Salpeter
IMF. Figure\,3 is confirming the results from Figures 1 and 2, and at the median
stellar mass of the distant galaxies (log($M_{\star}$)=10.55), their median
abundance is 0.4 dex lower than the local SDSS galaxies.  One can rule out that
this effect can be related to differences in the estimation of the stellar masses,
as it stands when considering absolute K band luminosities.

\begin{figure}
\centering
\input epsf
\epsfxsize 7.8cm
\epsfbox{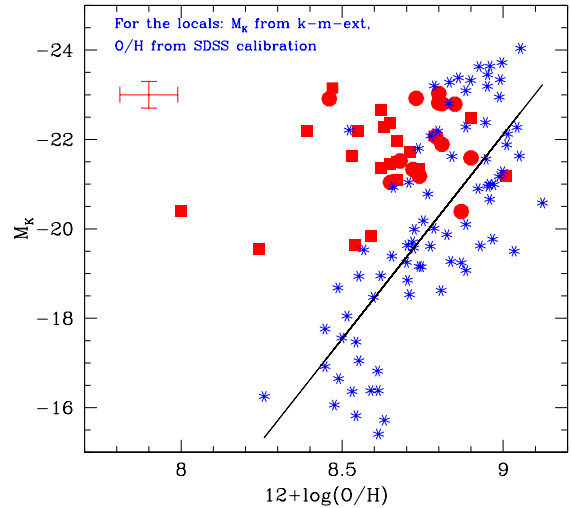}
\caption {$M_K$-metallicity relation for the intermediate-$z$ galaxies (red
symbols) and local spirals (blue stars) from the NFGS. The red filled circles
refer to distant LIRGs (SFR $>$ 20 $M_{\odot}$/yr)  and the red filled squares
refer to distant starburst galaxies (SFR $<$ 20 $M_{\odot}$/yr). Abundances
have been generated following the Tremonti et al. (2004) calibration for all
galaxies with typical uncertainty of 0.09\,dex, which has been plotted on the
left-top on the figure, toghether with the typical uncertainty of $M_K$
$\sim$0.3\,mag. Straight line is the best fit for the NFGS galaxies (y=-9.15x+60.23).
Because the intermediate-$z$ galaxies sample a much shallower range in $M_K$,
we won't fit them. Recall that these galaxies have been selected in I-band (or
B or V at rest), which leads us to suspect that some galaxies might have been
missed at $M_K$$>$$-$21. 
}
\label{fig1}
\end{figure}

\begin{figure}
\centering
\input epsf
\epsfxsize 7.8cm
\epsfbox{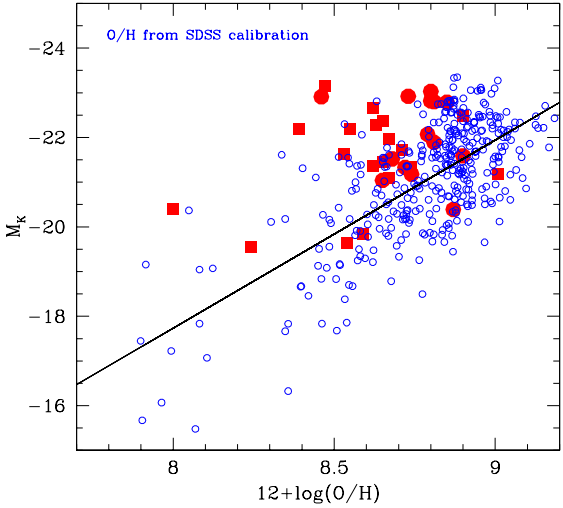}
\caption {$M_K$-metallicity relation for the intermediate-$z$ galaxies and for
the local KISS starbursts. Same symbols as in Fig.\,1 are used for distant galaxies,
and the local starbursts are represented by blue open circles. Abundances have been
generated following the Tremonti et al. (2004) calibration for all galaxies. 
The straight line is the best fit for the KISS
galaxies (y=-4.21x+15.93). 
}
\label{fig2}
\end{figure}

\section{Strong evolution of the stellar mass ($M_K$)-metallicity relation}

We have compared the metal abundance properties of a sample of 34 galaxies from
$z$=0.4 to $z$=1 (median redshift $z$=0.65) to those of nearby galaxies. We
carefully estimate oxygen abundances and stellar masses in a similar way for
both distant and nearby galaxies. It shows that the 34 distant galaxies show a
distribution that is systematically shifted towards lower abundances when
compared to local galaxies. Because the amplitude of the effect is much larger
than the error budget, it is tempting to claim that the stellar
mass-metallicity relation has strongly evolved since $z$$\sim$0.65, or in other
words, that intermediate mass galaxies at $z$$\sim$0.65 have on average half of
the metal content of local counterparts in their gaseous phase. Before trying
to derive firm conclusions from this result, let us investigate if some
systematics are involved.  

First, the emission line regions sampled by the slit (or fiber for the SDSS)
for local galaxies are not accounting for all the galaxy, unlike the case for
small and faint distant galaxies. 
However, the effect should be modest here. The emission line fluxes quoted by
Jansen et al. (2000a,b) have been obtained after scanning of the overall
galaxies in the NFGS (the integrated ones).  Tremonti et al. (2004) obtained
the Bayesian metallicity estimates by using the approach outlined by Charlot et
al. (2005), based on simultaneous fits of all the most prominent emission
lines, with a model  designed for the interpretation of integrated galaxy
spectra  (Charlot \& Longhetti 2001). They have checked the relations between
their Bayesian metallicity estimates and the $R_{23}$ values, which are
consistent with the relations from literature, including McGaugh (1991) and
Zaritsky et al. (1994).  You may worry about the 3{$^{\prime\prime}$} fiber
diameter of SDSS will focus more on  the nuclear part spectra of the galaxies,
and cause slightly higher metallicity estimates than the global ones. Indeed,
Kewley et al. (2005) has clearly discussed this aperture effect on the basis of
the NFGS spectra. They concluded, since the sample galaxies studied by Tremonti
et al. (2004) range in  the redshifts within 0.03$<z<$0.25, the aperture
effects are not significant in the metallicity estimates and the L-Z
relations.  Salzer et al. (2005) have argued that their metallicity estimates 
may be slightly  higher than the integrated values for the extended spirals due
to the abundance gradients. 
The reported evolution of the stellar mass-metallicity at $z$$\sim$0.65 is
observed after comparing to either NFGS, KISS or SDSS, so we believe that
aperture effects are likely modest.

Second, how our sample is representative of the distant galaxy population, and
could selection effects explain partly or fully our result? Most of the
galaxies have been selected from the CFRS and from the Ultra Deep Fields (see
Liang et al. 2004). Originally, this sample of $M_B<$-20 galaxies has been
defined to include a significant fraction of LIRGs. Indeed those constitute
38\% of the sample, while Hammer et al. (2005) estimated the fraction of LIRGs
to be 15\% in a sample of distant and intermediate mass galaxies. We compared
(see Figures 1 to 3) the populations of LIRGs and starbursts in our sample.
Applying a Kolmogorov-Smirnov test provides that their distributions in $M_K$
are similar, with a Student's parameter t=1.1 (the probability for the two
sub-samples drawn from a different population is 30\%). Moreover, LIRGs show on
average a higher O/H ratio than starbursts at a marginal level, with a
Student's parameter t=2.2 (the probability for the two sub-samples drawn from a
different population is 97\%). This marginal effect is expected because LIRGs
are producing metals and stars at a very efficient rate. So, introducing an
excessive number of LIRGs in our sample could only minimize the discrepancy
that we find with local samples.  

Third, we have to investigate if the effect is actually due to a deficiency of
metals at a given stellar mass (or $M_K$ magnitude) or if it could be related
to a brightening of the distant galaxies. At the median oxygen abundance
(12+log(O/H)=8.67) the brightening in K band magnitude would be
$\Delta$$M_K$=2.7 magnitudes (see Figure\,1), which should be compared to
$\Delta$$M_B$=3.1 for the same sample (see Figure\,4 of Hammer et al. 2005). 
Such a huge brightening at all wavelengths  is simply not plausible: it could
occur only if almost all the light in all distant emission line galaxies would
be associated with very young stars and strong dust. One has to remind that the
spectra of all intermediate mass galaxies show a mixture of young, intermediate
age and old stars, as evidenced by Balmer and metal (CaII H\,K, CaI, G-band,
MgI) absorption lines (see Hammer et al. 2005). We believe it is more natural
that the important population of distant emission line galaxies  (at $z$=0.65,
65\% of $M_B$$<$-20 galaxies have $W_0$(OII)$>$15\AA, see Hammer et al. 1997)
evolve at almost a constant luminosity into comparatively metal-rich disk
galaxies in the local universe, rather than to fade into dwarf galaxies. 

The closed box model for the evolution of galaxies (without gas inflow or outflow)
allows us the possibility to crudely estimate the evolution of the gaseous content.
Following Kobulnicky and Kewley (2004, see the original version in Searle \&
Sargent 1972 and Tinsley 1980), the metallicity Z is related to the gas mass fraction
$\mu$=$M_{gas}$/($M_{gas}$+$M_{star}$), by:
\newline Z=Y ln(1/$\mu$), 
where Y is the yield. By differentiating this relation,   we note that a change
in metallicity with gas mass fraction is independent of the yield and is given
by (see Kobulnicky \& Kewley 2004):
\newline d(logZ)/d$\mu$=0.434/($\mu$ln($\mu$)).
\newline In Figure\,3, we have plotted the closed box model evolution assuming the
median point of our sample (12+log(O/H)=8.67 and log($M_{\star}$)=10.55 at a
median redshift of 0.65), which links it to a local galaxy in the Tremonti et al.
(2004) sample with 12+log(O/H)=9.15 and log($M_{\star}$)=10.7. For the local
galaxy we assume a gas fraction $\mu$=10\%, which implies a median gas fraction at
$z$=0.65 of 30\%. In other words, in a closed box model, a present-day spiral would
have converted 20\% of its total mass from gas to stars during that redshift
interval. Assuming a higher gas mass fraction today ($\mu$=20\%), it would increase
the gas-to-star converted mass fraction to 25\% since $z$=0.65.   These values are
significantly higher than those found by Kobulnicky \& Kewley (2004), and such a
discrepancy is explained in the section below. 

\begin{figure}
\centering
\input epsf
\epsfverbosetrue
\epsfxsize 9cm
\epsfbox{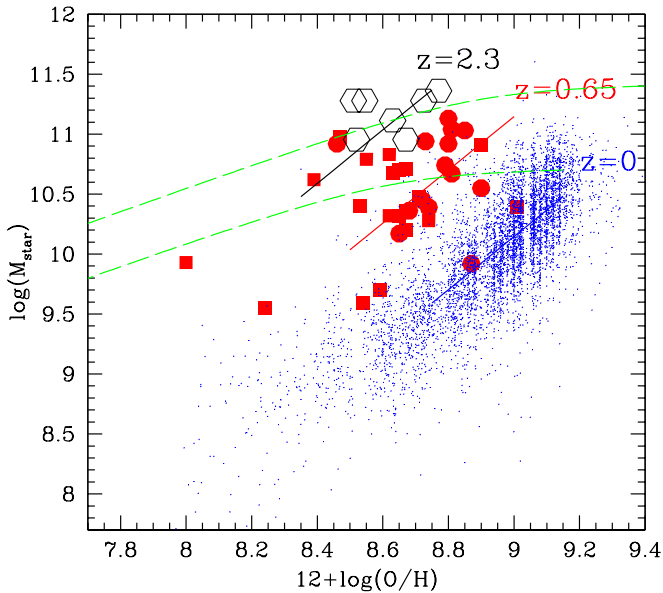}
\caption {Stellar mass-metallicity relation for the SDSS galaxies ($z$=0,
Tremonti et al. 2004, Bell et al. 2003, see text), the intermediate-$z$ 
galaxies ($z$$\sim$0.65, same symbols as in Figure 1) and the 7 galaxies (black
open hexagon) at $z$=2.3 described by Shapley et al. (2004).  Abundances have
been generated following the Tremonti et al. (2004) calibration for all
galaxies. Here we consider SDSS galaxies with  log($M_{star}$) $>$9.5 which are
fitted by a straight line with y=2.22x-9.83. This choice has been motivated by
the fact that all intermediate-$z$  galaxies have a stellar mass above this
cut-off. By forcing the lines to fit the median of intermediate distant
galaxies and of $z$=2.3 galaxies, we find an evolution of -0.4 dex and -0.7 dex
in metal abundances, respectively. The green long-dashed lines show the result
of a closed box model and indicating the general evolutionary trend from
$z$=2.3 and $z$=0.65 to $z$=0, respectively (see text).}
\label{fig3}
\end{figure}

\subsection{Comparison with other studies of intermediate-$z$ galaxies}

 There is only one study which can be directly compared to our result, namely,
the paper presented by Savaglio et al. (2005), in the frame of the Gemini Deep
Deep Survey (GDDS). Their method (flux calibration, removing of underlying
absorption etc...) is essentially similar to ours. Their sample consists of 24
0.4 $<$ $z$ $<$ 1 galaxies with only 4 of them having log($M_{star}$) $>$ 10,
i.e. GDSS galaxies are far less massive than those presented here. They found
an average extinction of $A_V$=2.2 for their sample, a value similar to that
found by Liang et al. (2004) but significantly higher than $A_V$=1, a value
assumed by Lilly et al (2003). Notice that one third of the Lilly et al. (2003)
sample is made of LIRGs and that the inferred metallicity assuming $A_V$=1
leads to values $\sim$0.3 dex larger than our estimates based on a proper
correction of the \ion{[O}{ii]}$\lambda$3727/$H\beta$ ratio. Indeed Savaglio et
al. (2005) also find a significant evolution of the mass-metallicity relation.
However, comparison to their results is complicated by the fact that they have
merged their sample (low masses, on which they assumed $A_V$=2.2)  to that of
Lilly et al. (high masses, on which they assumed $A_V$=1). Recall that for
starbursts, a general correlation is expected between mass, metallicity, dust
and then extinction which contradicts the Savaglio et al (2005) assumption.
This is particularly true at intermediate redshift and the emergence of
relatively massive LIRGs at $z$$>$ 0.4 (Hammer et al. 2005) is evidencing the
serious impact of extinction effects in estimating the evolution of the stellar
mass-metallicity relationship. 

 Kobulnicky \& Kewley (2004) have described the metal-$M_B$ relation of 204
galaxies in the GOODS-North field, and found a much more moderate evolution
than us. By comparing their result to that of Liang et al. (2004) or Hammer et
al. (2005), they found at a constant $M_B$ a decrease in metal abundance of
0.14 dex from $z$=0 to 1, which compares to 0.3 dex from $z$=0 to $z$=0.7 in
Hammer et al. (2005). Because there are no evidence indicating any serious
difference in the selection of galaxies between the various studies, we
investigate if this could be due to the different methodologies adopted here
and there. Recall that Kobulnicky \& Kewley (2004) have to derive $R_{23}$
parameters from equivalent width ratios, because they are using non flux
calibrated spectra. Their method has been calibrated from systematic comparison
to the KISS and NFGS samples (see Kobulnicky \& Philips 2003).

 To do so, we have applied for our sample the equivalent width methodology, and
Figure 4 compares the values derived by both methods. Notice that the
equivalent width method producing systematically higher metal abundance by 
$\sim$ 0.2 dex, which is likely to be the origin of the discrepancy. Indeed the
discrepancy is even higher for large masses and large extinction coefficients: 
Kobulnicky and Kewley (2004) were finding an absence of evolution  at the high
mass end, which is not confirmed by  our study. We suspect that the  Kobulnicky
and Phillips calibration might not apply to distant starbursts because those
show on average, larger extinction coefficient than local galaxies.

\begin{figure}
\centering
\input epsf
\epsfverbosetrue
\epsfxsize 9cm
\epsfbox{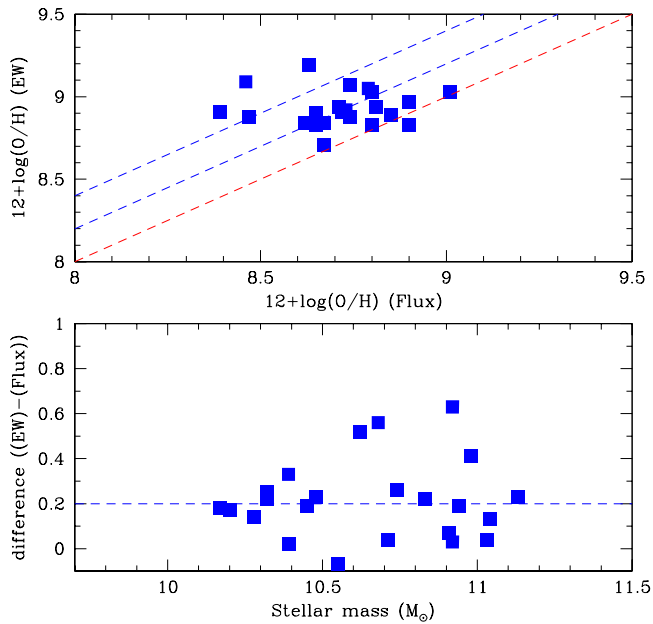}
\caption {{\it Top:} Comparison of metal abundances derived from extinction
corrected flux ratio and those derived following the equivalent width method as it
is described in Kobulnicky and Kewley (2004); {\it bottom:} shows the difference
of oxygen abundances from the two methods against the stellar masses.
}
\label{fig4}
\end{figure}

\subsection{Comparison to higher redshift galaxies}

Shapley et al. (2004) presented the near IR spectroscopic measurements for
seven star-forming galaxies at $2.1<z<2.5$.  The oxygen  abundances of these
galaxies were estimated  by using \ion{[N}{ii]}/H$\alpha$ ratio following
calibrations from  Pettini \& Pagel (2004) and Denicolo et al. (2002), and show
12+log(O/H)=8.47-8.69 and 8.56-8.80, respectively, which are $\sim$0.3-0.4\,dex
lower than solar values. Shapley et al. (2004) found that these galaxies
already possess stellar masses in excess of 10$^{11}M_\odot$.  

Figure 3 presents the variance of the stellar mass - metallicity relation at
$z$=0, 0.65 and 2.3. At $z$=0.65, galaxies with log($M_{star}$) $>$10.5  or
with 10$<$log($M_{star}$) $<$10.5 are metal deficient when compared to SDSS
galaxies: from the small data set presented here, there is no evidence for an
evolution of the slope of the stellar mass-metallicity relationship.  Because 
at $z$=0.65 (and more obviously at $z$=2.3) only high stellar masses are
included, we therefore consider in the following an absence of slope
evolution.  To fit $z$=0.65 and $z$=2.3 galaxies, it is required to shift the
local relation by -0.4 and -0.7 dex in metal abundance, respectively.  Assuming
a closed box model, would lead to median gas fractions of 30\% and 45\% at 
$z$=0.65 and $z$=2.3, respectively. In the absence of knowledge of the
evolution of metal abundance for $z$=2.3 intermediate mass galaxies, it is
difficult to translate these values directly to average mass formation within
the $z$=0.65-2.3 interval. Nevertheless, combining of our result with that of
Savaglio et al. (2005) suggests a formation of 1/5 to 1/4 of the stellar mass
in most present day spirals since $z$=0.65, which agrees rather well with
expectations from stellar mass density studies (see Drory et al. 2004). Recall
that it matches well with the 40\% of stellar mass formed since the last 8 Gyr
claimed by Hammer et al. (2005) from integrations of the IR star formation rate
density.

\section{Conclusion: gas converted to stars since the last 8 Gyr }

There is a growing consensus that a significant part of the masses of the 
present-day spirals have been formed since $z$=1 (Hammer et al. 2005; Bell et
al. 2005).  According to Drory et al. (2004, see their Figure 7), the stellar
mass density at $z$$\sim$0.65 is roughly 75\% that at $z$=0. This result
matches well with the integrated star formation density, if IR light is taken
into account. Here we have shown that it also matches with the evolution of the
gaseous fraction in galaxies, which leads, under our assumption of a closed box
model, to 20-25\% of the gas which has been transformed into stars since
$z$=0.65. Of course, a closed box model is a very crude approximation of the
evolution of intermediate mass galaxies, which can be affected by gas outflow,
inflow and minor or major mergers. However, as a whole, intermediate mass
galaxy population can be assumed to characterize most of the gas to mass ratio
of the universe, and applying a closed box  model to this population allows
comparison to other integrated quantities such as stellar mass or star
formation densities. 

We find no evidence for a change in the slope of the  M$_{star}$-Z relation.
Notice that many studies (Lilly et al. 2003; Kobulnicky \& Kewley, 2004) found
no evolution for the higher mass range of their sample, which has been
interpreted as an evidence for the ``downsizing" scenario. While our data are
not that different from those of the above studies, we do find an evidence
that  intermediate mass galaxies show under-abundances when compared to local
galaxies. Our method accounts for a proper correction for extinction, that
being derived from both IR/optical ratio and Balmer line ratio.  Recall that
extinction potentially affects more the metal abundance estimates of massive
starbursts, and that it may lead to erroneous conclusions when its effect is
neglected. In a forthcoming paper (Flores, Proust and Hammer, in preparation),
we will compare our oxygen abundances to those from the \ion{[N}{ii]}/$H\alpha$
ratio which is essentially independent of extinction. 

There are also many recent publications to discuss the assembly history of
galaxies. It has been shown that the specific star formation rate
(SFR/M$_{star}$) is anti-correlated with the stellar mass in the local Universe
(see Brinchmann et al. 2004), and the b-parameter ($SFR/<SFR>$) on average
decreases with mass for the late-type galaxies (Salim et al. 2005). At higher
redshift, a similar trend was found in  several studies which were using
various SFR tracers from UV to mid-IR (Bauer et al. 2005; Feulner et al. 2005a,b;
Perez-Gonzalez et al. 2005). All these studies suggest a ``downsizing", in the
way that, on average, massive galaxies have lower specific star formation rates
than those with lower mass. Papovich et al. (2004) and Bundy et al. (2005)
obtained the similar results. Indeed, the metallicities of galaxies must be an
excellent indicator to explore this issue, because it links two independent
quantities (M$_{star}$-O/H). We notice however that this study cannot be
conclusive for or against the ``downsizing", because the present sample
includes too few objects in a too restricted mass range. In the near future we
will be able to better test this  by including a much larger set of
measurements ($\sim$ 200 intermediate redshift galaxies), in the frame of a
large program at VLT.

\section*{Acknowledgments}
  We thank the referee for the very valuable comments which have helped us
 in improving this paper.  We thank Dr. Rob Kennicutt for his very valuable
 comments in science and his warm help in improving the English description. 
 We thank Dr. John Salzer for providing his data in electronic format, and
 thank Dr. Chang Ruixiang and Shen Shiying for the discussion about the  SDSS
 database.  We are especially grateful to Dr. Ravikumar who kindly provides us
 the K band measurements for six additional objects to our previous sample.  We
 thank Prof. Chou Chih-Kang for his help in improving the English language in the
 text. We are grateful to Dr. David Crampton and Dr. Sandra Savaglio for enjoyable
 discussions and for communicating with us about their manuscript prior to
 publication. 
 We thank Dr. Lisa Kewley for the interesting discussion about aperture effects
 on metallicities of galaxies, also for her important comments on the results.
 This work was supported by the Natural Science Foundation of
 China (NSFC) Foundation under No.10403006 and No.10433010.

\newpage

\begin{table*}
\caption {The basic parameters of the sample of distant galaxies  for stellar
mass-metallicity relation: only the galaxies with log(M/M$_{\odot}$)$>$9.5. ``997"
means the line is blended with sky; ``998" means no corresponding line detected at
the line position; ``999" means the line is shifted outside the rest-frame
wavelength range. The last column gives more notes for several objects, `` LINER"
means the  Low Ionization Nuclear Emission line Region classified by following the
[OII]/H$\beta$ vs. [OIII]/H$\beta$ relation. }
{\scriptsize
 \begin{tabular}{cccccrrrccc} \hline
CFRS    & $z$ &  M$_{K}$ &log(M/M$_{\odot}$)& 12+log(O/H) & $R_{23}$ & [OII] & [OIII] & H$\beta$ & Note \\ \hline
00.0137 &  0.9512 &   -23.15 &  10.98  &   8.47$\pm$0.13   & Yes &     &     &     &   \\
00.0141 &  0.4401 &   -21.49 &  10.36  &       8.67$\pm$0.10   & No  & 999 &     &     &   \\
00.0564 &  0.6105 &   -21.61 &  10.44  &    ---       & No  & 999 & 998 &     &  \\
00.1721 &  0.5581 &   -21.09 &  10.20  &       8.67$\pm$0.05  & Yes &     &     &     &  \\ \hline
03.0006 &  0.8836 &   -19.59 &   9.58  &    ---        & No  &     & 999 & 999 &  \\ 
03.0035 &  0.8804 &   -23.11 &  11.07  &    ---       & No  &     & 999 & 999 &   \\
03.0062 &  0.8252 &   -22.68 &  10.87  &    ---       & No  &     & 999 & 999 & \\
03.0085 &  0.6100 &   -20.80 &  10.10  &    ---       & No  &     &     & 997 &  \\
03.0174 &  0.5250 &   -20.87 &  10.16  &    ---       & Yes &     &     &     & LINER \\
03.0186 &  0.5220 &   -21.95 &  10.61  &    ---       & Yes &     &     &     & LINER    \\
03.0422 &  0.7150 &   -22.47 &  10.76  &    ---       & No  &     & 999 &     &   \\
03.0442 &  0.4781 &   -20.40 &   9.93  &   8.05$\pm$0.30   & Yes &     &     &     &   \\
03.0445 &  0.5300 &   -22.07 &  10.74  &   8.79$\pm$0.07   & Yes &     &     &     &   \\
03.0507 &  0.4660 &   -21.04 &  10.17  &   8.65$\pm$0.05   & Yes &     &     &     &  \\
03.0523 &  0.6508 &   -21.52 &  10.36  &   8.68$\pm$0.09   & Yes &     &     &     &  \\
03.0533 &  0.8290 &   -22.41 &  10.73  &    ---           & No  &     & 999 & 999 &  \\
03.0570 &  0.6480 &   -20.39 &   9.92  &   8.87$\pm$0.27   & Yes &     &     &     &  \\
03.0589 &  0.7160 &   -21.08 &  10.21  &    ---           & No  &     & 999 & 997 &  \\
03.0595 &  0.6044 &   -21.44 &  10.35  &    ---           & No  &     &     &     & no A$_V$ \\
03.0645 &  0.5275 &   -21.34 &  10.28  &   8.74$\pm$0.09   & Yes &     &     &     &  \\
03.0776 &  0.8830 &   -20.65 &  10.02  &    ---           & Yes &     &     &     & LINER   \\
03.0932 &  0.6478 &   -22.82 &  10.92  &  8.80$\pm$0.19    & Yes &     &     &     &  \\
03.1309 &  0.6170 &   -22.91 &  10.92  &  8.46$\pm$0.11    & Yes &     &     &     & \\
03.1349 &  0.6155 &   -22.92 &  10.94  &  8.73$\pm$0.06    & Yes &     &     &     & \\
03.1531 &  0.7148 &   -22.28 &  10.72  &    ---       & No  &     & 998 & 997 &   \\
03.1540 &  0.6898 &   -22.31 &  10.70  &    ---       & No  & 997 &     &     &   \\
03.1541 &  0.6895 &   -21.18 &  10.39  &  8.74$\pm$0.09    & Yes &     &     &     & \\  \hline
22.0242 &  0.8638 &  -22.27  &  10.68  &        8.63$\pm$0.26  & Yes &     &     &     &             \\
22.0344 &  0.5168 &  -19.84  &   9.70  &        8.59$\pm$0.06  & Yes &     &     &     &   \\
22.0429 &  0.6243 &  -21.72  &  10.48  &    8.71$\pm$0.15  & Yes &     &     &     &    \\ 
22.0576 &  0.8905 &  -21.46  &  10.32  &    8.65$\pm$0.14  & Yes &     &     &     &    \\
22.0599 &  0.8854 &  -22.19  &  10.62  &        8.39$\pm$0.06  & Yes &     &     &     &    \\ 
22.0626 &  0.5150 &  -19.64  &   9.59  &        8.54$\pm$0.09  & No  & 999 &     &     &    \\
22.0637 &  0.5419 &  -21.37  &  10.32  &    8.62$\pm$0.08  & Yes &     &     &     &     \\       
22.0721 &  0.4070 &  -20.46  &   9.99  &      ---     & No  & 999 &     &     &     \\
22.0779 &  0.9252 &  -22.37  &  10.70  &        8.65$\pm$0.12   & Yes &     &     &     &     \\
22.0828 &  0.4070 &  -20.53  &  10.01  &      ---     & No  & 999 &     &     &    \\
22.0919 &  0.4714 &  -19.55  &   9.55  &    8.24$\pm$0.03  & Yes &     &     &     &     \\
22.1064 &  0.5369 &  -21.64  &  10.40  &    8.53$\pm$0.07  & Yes &     &     &     &         \\
22.1070 &  0.8796 &  -22.65  &  10.83  &        8.62$\pm$0.27  & Yes &     &     &     &    \\     \hline  
UDSR    &         &          &         &               &       &   &   &   \\
  08    & 0.7291 &  -21.33   &  10.45  &     8.72$\pm$0.04  & Yes   &   &   &   \\
  10    & 0.6798 &  -23.03   &  11.13  &     8.80$\pm$0.07  & Yes   &   &   &   \\
  14    & 0.8150 &  -22.79   &  11.03  &     8.85$\pm$0.26  & Yes   &   &   &   \\
  15    & 0.4949 &  -21.89   &  10.67  &     8.81$\pm$0.05  &  No   &  999 &   &   \\
  23    & 0.7094 &  -22.80   &  11.04  &     8.81$\pm$0.05  & Yes   &   &   &   \\
  26    & 0.3841 &  -22.19   &  10.79  &     8.55$\pm$0.24  & Yes   &   &   &   \\ \hline
UDSF    &        &           &         &                    &       &   &   &   \\
  02    & 0.7781 &  -21.97   &  10.71  &     8.67$\pm$0.49  & Yes   &   &   &   \\
  03    & 0.5532 &  -21.59   &  10.55  &     8.90$\pm$0.07  & Yes   &   &   &   \\
  07    & 0.7014 &  -22.48   &  10.91  &     8.90$\pm$0.10  & Yes   &   &   &   \\ 
  16    & 0.4548 &  -21.19   &  10.39  &     9.01$\pm$0.04  & Yes   &   &   &   \\ \hline \hline
\end{tabular}
  }
\end{table*}

\end{document}